# Formation of extended thermal etch pits on annealed Ge wafers

L. Persichetti,[1,*] M. Fanfoni,[2] M. De Seta,[1] L. Di Gaspare,[1] L. Ottaviano,[3] C. Goletti,[2] and A. Sgarlata[2]

[1]*Dipartimento di Scienze, Università Roma Tre, Viale G. Marconi, 446- 00146 Roma, Italy*

[2]*Dipartimento di Fisica, Università di Roma "Tor Vergata", Via Della Ricerca Scientifica, 1- 00133 Roma, Italy*

[3]*Dipartimento di Scienze Fisiche e Chimiche, Università degli Studi dell'Aquila, L'Aquila, Italy*

An extended formation of faceted pit-like defects on Ge(001) and Ge(111) wafers was obtained by thermal cycles to $T > 750$ °C. This temperature range is relevant in many surface-preparation recipes of the Ge surface. The density of the defects depends on the temperature reached, the number of annealing cycles performed and correlates to the surface-energy stability of the specific crystal orientation. We propose that the pits were formed by preferential desorption from the strained regions around dislocation pile-ups. Indeed, the morphology of the pits was the same as that observed for preferential chemical etching of dislocations while the spatial distribution of the pits was clearly non-Poissonian in line with mutual interactions between the core dislocations.

[*]luca.persichetti@uniroma3.it

## I. INTRODUCTION

Despite recent advances in the van der Waals epitaxy of two-dimensional semiconductors [1-7], bulk group-IV still plays a leading role in current complementary-metal-oxide-semiconductor (CMOS) technology. Within group IV, non-silicon-oriented research is mainly devoted to Ge. SiGe heterostructures monolithically grown on Si are the ideal test bed for understanding alloying [8-11] and the interplay between elastic/plastic relaxation at the nanoscale [12-22] Their also being promising candidates for integrating optical communications into a CMOS platform, thanks to their optical properties potentially compatible with the C-band transmission window [23]. Ge wafers, on the other hand, are the substrates of choice for the epitaxial growth of high-efficiency multi-junction solar cells based on III-V semiconductors [24-29] and have also been shown to be suitable CMOS compatible templates for graphene overgrowth [30-32]. All these applications require a highly-demanding surface quality of the epi-ready Ge substrates. Indeed, any deviation from a perfect surface will be a nucleation point for a defect or a cause of a non-uniform epi-stack. In particular, one of the main issues affecting the manufacturing of semiconductor wafers is the formation of extended secondary defects such as crystal-originated pits (COPs) and L-pits (also referred to as A-Swirls) resulting from the aggregation of intrinsic point defects [33]. While COPs are voids produced by the aggregation of vacancies, it is generally assumed that the formation of L-pits is related to dislocation loops either intrinsically formed during the wafer manufacturing [34, 35] or by misfit strain in the case of SiGe heteroepitaxy [36]. In the former case, possible sources of dislocation loops are the polishing treatment and, above all, the precipitation and agglomeration of self-interstitials oxygen or carbon trapped during the wafer pulling [37, 38]. The aggregation of interstitials exerts a sizeable amount of stress in the wafer matrix which is finally released with the generation of a network of dislocation loops around the original defect [33]. In contrast to COPs which have a size typically below 100 nm and can be annealed out, L-pits are micron in size and extremely stable against thermal curing methods. Therefore, while COPs impact on the device performance only if the location of a void coincides with that of an active element, damage caused by L-pits, due to their size, is always permanent. In addition, L pits at the substrate are too large to be buried within an epilayer film which will be also damaged by their presence. Consequently, these imperfections are detrimental for the final performance and the production yield of any device monolithically grown on Ge substrates. In addition, they can be nucleation sites for fracture, thus affecting the mechanical strength and the rate of early failures of devices due to wafer breakage. Mechanical robustness under high-stress conditions is particularly important for the specific market of Ge wafers in which space applications of triple-junction solar cells on Ge substrates is the major component [25]. Since L-pits are totally unacceptable to device manufactures, much effort has gone into reducing their impact on Si wafer production [33, 37, 39]. Conversely, much less information is available for Ge substrates. In this paper, we performed a detailed investigation of pit-like defects in Ge wafers, correlating their density to specific treatments of the surface before epilayer overgrowth. We found that specific thermal treatments of the Ge surface result in faceted micrometer-size pits in wafers which are initially free of visible defects. We suggest that the pits decorate dislocation pile-ups formed by thermal stress during repeated annealing cycles, starting from initially-silent aggregations of point defects embedded in the wafer, like L-pits and large voids, which reach the critical lattice strain for dislocation introduction. Notably, the temperature needed for the formation of pits was comparable to that used for ultra-high-vacuum (UHV) surface cleaning of Ge [40, 41] and for the graphene/Ge growth [30-32]. The pits were morphologically very similar to those produced by preferential chemically etching of dislocations [42, 43]. The density of the defects depends on the maximum temperature reached, the number of thermal cycles performed and, above all, on the wafer orientation. By statistically analyzing the distribution of the pits, we show evidence for ordering at short-range compatible with mutual elastic interactions between dislocations. Locally, the pits demonstrate wall patterns typical of dislocation-pinning on swirl defects.

**II. MATERIALS AND METHODS**

Experiments were carried out by using commercial epi-ready, prime-grade polished Czochralski (CZ) Ge wafers (N-type, Sb-doped and resistivity of about 5-7 $\Omega$ cm) with (001) and (111) orientations. The dislocation density of the crystals determined by the manufacturer is ~$10^3$ cm$^{-2}$. Annealing treatments were performed in UHV, using direct-current heating. The temperature was measured using an optical pyrometer and calibrated by determining the melting temperature on a sacrificial sample. Characterization was performed *ex-situ* by atomic force microscopy (AFM) operated in tapping mode (Veeco Nanoscope IIIa MultiMode), field emission-scanning electron microscopy (FE-SEM, LEO 1550) and optical microscopy (Nikon L-IM equipped with a digital camera DS-5M). Optical images were processed using background-subtraction, thresholding and object-counting algorithms available in the ImageJ package. AFM images are processed with the WSxM software.

## III. RESULTS AND DISCUSSION

*Annealing-induced pit-like defects on Ge(001) and Ge(111)*

UHV cleaning of Ge substrates by ion sputtering or wet chemical etching typically requires the repetition of several annealing steps where, in most cases, a peak temperature between 750 °C and 850 °C is reached [40, 41, 44]. The focus of the reports is generally on the quality of the surface at the nanoscale, whereas the mesoscale morphology was ignored. Nonetheless, for any device-oriented application, the evaluation of the surface quality has to be extended to a much larger length scale, at least of several hundreds of microns. Here, we found that annealing of Ge substrates above 750° may be detrimental for the quality at the mesoscale, resulting in the formation of an extended aggregation of pit-like defects at the surface. The standard protocol used for annealing consists of 7 thermal cycles in which the temperature is ramped up from room temperature to $T_{max}$ with a ramp rate of about 150 °C/min, maintained at $T_{max}$ for 20 min and, subsequently lowered again to room temperature using the same rate. $T_{max}$ was varied between 750 and 850 °C. The ramp rate for cooling is slow enough to avoid irreversible roughening of the Ge(001) surface [45, 46]. Thermal cycles mimic the repetition of annealing steps during UHV cleaning of Ge substrates by ion sputtering or wet-etching.

Figure 1 shows optical microscopy images of the Ge(001) surface after 7 thermal cycles to (a) 850 °C and (b) 810 °C, compared to the surface before annealing (c). The formation of pits after annealing is evident. The density of the pits increased with the annealing temperature, as shown in Table 1. At a given temperature, we found that the density of the pits increased with the number of annealing cycles performed, reaching a saturation value for 7 thermal cycles. A larger number of cycles did not increase the density further. Conversely, for less than two annealing cycles, the number of pits observed was negligible. On a subset of the samples, we performed a standard sputtering/annealing cleaning prior to the final annealing cycle which did not impact on the defects' density.

Higher optical magnification [inset of panel (a)] reveals that most of the pits have the shape of an inverted pyramid with a square base and edges oriented along the <110> directions. A better view is obtained by SEM (Fig. 2) which also shows that some of the pits are elongated along the <110>. From the SEM images, it is clear that the pits are perfectly faceted. The AFM investigation [Fig. 3(a,b)] confirms that the exposed facets have a preferential orientation compatible with that of {111} facets. Quantitative information on faceting are obtained by applying the so-called facet plot (FP) analysis [47]. It consists of a 2D diagram in which the position of each spot represents the local normal orientation relative to the substrate plane, which corresponds to the center of the diagram, while the intensity shows the relative amount of the surface with that orientation. The four-fold symmetric spots in the FP shown in Fig. 3(b) thus indicate that the surface breaks up into four crystallographic {111} faces. Since the (111) face is the most stable orientation for Ge [48], it is reasonable to assume that surface energy has a major role in stabilizing the morphology of the pits. An interesting comparison is therefore with the Ge(111) substrate. Figure 4(a) shows a large-scale optical image of the Ge(111) surface after 7 annealing cycles to 820 °C. One can notice the appearance of micrometer-size pits which, in contrast to those observed on Ge(001), have a triangular symmetry [Fig. 4(b)]. The three-dimensional topography of the pits obtained from AFM images [Fig. 3(c,d)] is in fact that of a three-fold symmetric inverted pyramid bounded by {001} facets. Accordingly, three spots separated by 120° appear in the corresponding FP shown in Fig. 3(d). Less than 10% of the pits has a flat bottom face [Fig. 4(b) and inset of Fig. 3(d)], corresponding to a stacking fault of the (111) surface. It is interesting to note that the different symmetry and morphology of pits formed on Ge(001) and on Ge(111) can be explained in terms of geometry. In the cubic-*m3m* point-group symmetry of Ge crystals, the equilibrium Wulff shape minimizing the total surface energy is a truncated octahedron, when the set of facets observed in the experiment is considered [i.e. {001} and {111}]. As shown in Fig. 3(e), for a given {001} face, there are four {111} facets and four <110> crystal edges. These four {111} facets define a four-fold square-based pyramid with the (001) plane. Conversely, for a given {111} face, there are only three {001} facets [Fig. 3(f)]. It is easy to see that these three {001} facets define, instead, a three-fold {100} pyramid with a triangular base.

From Table 1, one can see that the density of pits is markedly higher on the Ge(001) substrate, whereas the pits on the Ge(111) surface are, on average, larger. In order to directly compare the stability of the {001} and the {111} faces against the formation of pits, we evaluated the fraction of the surface covered by pits as a function of temperature by analyzing the optical microscopy images with a thresholding algorithm using the ImageJ software. Figure 3(g) shows the scaling with temperature of the fraction of surface covered by pits on Ge(001) and on Ge(111). It is evident that the Ge(001) surface is much more strongly affected by the formation of pits. We believe that this strong impact of the wafer orientation reflects the significant orientational anisotropy of the surface energy of Ge. Since the Ge(111) face is in fact considerably more stable than the Ge(001) one [48], the growth of pits on the (111) substrate has a larger cost in terms of surface energy. Conversely, the energy cost of pitting due to the formation of additional exposed surfaces is diminished, in part, on Ge(001) by faceting along the {111} minimum of the surface energy.

It is noteworthy that the pits observed here have precisely the same shapes (i.e. square for the (001) and triangular for the (111) substrate) as those produced by preferential chemical etching of dislocations on Ge(001) and Ge(111) [42, 43]. Therefore, we expect that, also in the present case, dislocations play a critical role in the formation of pits. In preferential chemical etchings, the region around where a dislocation emerges from a surface is more vulnerable to chemical attacks, due to the higher local strain, and it is therefore preferentially etched away, leading to the formation of a pit. In close analogy, we believe that, in our case, the pits were produced by the preferential removal of material from the highly-strained region around a dislocation or a dislocation pile-up, although the process was thermally rather than chemically activated. In the high-temperature range where the formation of pits is observed, the mobility of Ge atoms was high enough to activate long-range surface diffusion. In the presence of strain, a dominant contribution to the diffusional field arises due to the gradient of the strain energy density $\Xi$, $J_\Xi \sim -\nabla \Xi$ [49]. Due to this term, material was transferred from high-strain regions towards more relaxed areas. This means, in our case, a net flux which depleted the strained areas around the dislocation pile-ups, leading to the formation of pits.

Since the density of the pits observed after thermal treatments is at least two orders of magnitude higher than the initial density of dislocations in the wafers, we believe that thermal stress originated during temperature cycles contributes decisively to generation and multiplication of dislocations. Moreover, the observation that pits do not appear after a single annealing step shows that it is the entire thermal history of the sample which matters, thus indicating a cumulative effect of stress generated by temperature gradients throughout multiple annealing cycles. Compared to silicon, the higher density of germanium combined with its lower thermal conductivity and mechanical strength, enhances the amount of thermal stress generated and, in turn, determines a lower critical resolved shear stress [50]. At the same time, the mobility of dislocations in germanium is about two orders of magnitude higher than in silicon [51]. All together, these observations suggest that additional dislocations can be generated by thermal stress during repeated annealing treatments of germanium and these dislocations, once produced, have enough mobility to reach the surface and form pile-ups. The generation of dislocations may be initiated at pre-existent defects in the wafer, such as L-pits and large voids [33], where a strain in the lattice is already present due to stacking faults originating from the aggregation of self-interstitials oxygen or carbon impurities trapped during the wafer pulling [37, 38]. Silent defects embedded in the as-grown wafers, for which the strain field is initially below the threshold for dislocation generation, may be activated by repeated annealing cycles to $T > 750\ °C$. Since the concentration of this kind of defects become progressively larger as the number of dislocations in the manufacturing of CZ wafers is reduced [52], the detrimental results of repeated annealing cycles should be considered for designing optimal surface preparation recipes of Ge substrates.

*Spatial equilibrium distribution of pit-like defects*

Consistently with the mechanism above, we found that the spatial distribution and local ordering of pit-like defects shared interesting similarities with dislocation patterning. The self-organization of dislocation structures produced by rearrangement under mutual interaction is characterized by frustration between forming locally ordered structures and randomization of the spatial distribution at larger scale [53]. This is the result of the interplay between long-range elastic repulsion via Peach-Koehler forces that decay as $1/R$ (where $R$ is the separation between points on dislocation loops) and topological attractive interactions as in the case when two dislocation segments interact to form a dipole [54]. The fingerprint for such a complex interaction scenario is the spontaneous formation of alternating dislocation-rich and dislocation-poor regions, as well as the building up of tangles and walls when the dislocations were trapped and immobilized, for example, at local pinning defects [55]. The formation of walls is indeed the most striking pattern of pits observed on our samples (Fig. 5). Typically, the walls cross the sample along its whole length [Fig. 5(a)] and are composed either by individual pits aligned along the wall direction or by a tangle of aligned pits [Fig. 5(b)]. We believe that the walls are the result of dislocations pinned at defects of the wafer. In the case of the longest walls, the pinning sites are probably extended swirl defects which typically spread over the whole wafer surface. Away from the walls, a possible spontaneous organization of the pits into non-random spatial distribution cannot be visually detected. A powerful descriptor for a deviation from randomness in a two-dimensional (2D) distribution of points is the Voronoi tessellation [56, 57]. It is known that a random generation of points in a 2D space results in a tesseral distribution of Voronoi cells following a standardized Gamma function

$$\gamma_\alpha(x) = \frac{\alpha^\alpha}{\Gamma(\alpha)} x^{\alpha-1} e^{-\alpha x} \qquad (\text{Eq. 1})$$

with $\alpha = 3.5\text{-}3.6$ [56]. As a degree of correlation is gradually introduced, the suitability of this distribution in fitting the distribution of the Voronoi cell areas progressively vanishes. Figure 6(a) shows the tesseral distribution of Voronoi cells obtained from optical microscopy images for 7 annealing cycles to 850 °C on the Ge(001) substrate. At the magnification used, the actual shape and size of the pits could be neglected and, thus, the point-like approximation was used, considering the centers of mass of the pits. In order to minimize the boundary effects, the cells close to the edges of the image were not taken into account for distribution. It is evident that the $\gamma_{\alpha=3.6}$ function in Eq. 1, displayed as a continuous line in the plot, totally fails to describe the distribution of the Voronoi cell areas obtained from the experimental image (dots). In particular, we noticed that *(i)* the $\gamma_{\alpha=3.6}$ function, corresponding to randomly-generated points, has a completely different behavior at small cell areas, going to zero less abruptly than in the experiment; *(ii)* the experimental distribution was markedly narrower and less asymmetric than that expected for a Poisson process (i.e. for the experimental distribution, the variance is about half of that obtained for $\gamma_{\alpha=3.6}$). Both these observations hint at the presence of a certain degree of correlation among the pits. In fact, *(i)* the behavior at small cell areas is consistent with the existence of an exclusion zone around each pit which reduces the abundance of small cell areas compared to the Poissonian case and *(ii)* within the limit of a perfectly ordered configuration, a $\delta$-like tesseral distribution is expected. Physically, non-randomness is in line with a spatial distribution biased by mutual interactions between the dislocations decorated by pits during high-temperature annealing cycles. In order to obtain further insight into the degree of order effectively present, we evaluated a second major descriptor, known as integrated conditional density, which represents the behavior of the density of pits at various length-scales from the center of each pit [58]. Operationally, we considered a given pit and a circle of radius *r* centered on it. We then computed the density in the circle, by counting all the pits that lie within the circle, normalizing the number to the area of the circle itself. We repeated the operation for all the pits in the image. The result obtained is shown in Fig. 6(b) as a continuous curve. In the limit of large *r*, the asymptotic value in the plot is obviously the nominal density in the image. We compared the experimental curve to what is found for a Poissonian (dashed curve) and an ordered square-array configuration of points (dotted curve) having the same asymptotic density as the test ensemble. It is clear that, while the pit distribution does not show the multiple-peak structures characteristic of long-range order, a residual degree of correlation is nevertheless present in the pattern, compared to a random generation of points. With respect to the Poisson case, the experimental density saturates to the asymptotic value much more gradually. In particular, the density at short distance is lower than in the case of a random ensemble. This is consistent with the existence of an exclusion area around each pit, well matching the Voronoi analysis. At higher separations, the density first increases to a value larger than the asymptotic one and, then, saturates towards the asymptotic plateau. We believe that this behavior reflects the tendency to spontaneously cluster into areas (being several hundred of micrometers in size) where the density of the pits (and, in turn, of the core dislocations) is higher. Evidence for clustering in the spatial pattern of pits was gained by dividing the test image into quadrants and counting the number of pits falling in each of these sub-regions. As a density function, we used the quartic kernel

$$\hat{\lambda}_h(\mathbf{r}) = \sum_{d_i \leq h} 3\pi^{-1} h^{-2} \left(1 - d_i^2 h^{-2}\right)^2$$, where $d_i$ is the distance between a point located at *r* in the study region and a pit at $r_i$,

while *h* is the quadrant size. An example is shown in Fig. 7(a) where the density function $\hat{\lambda}_h(\mathbf{r})$ across the image is visualized using an inverted gray-scale color map. As evident, the spatial distribution of pits is non-homogeneous, since dark areas where the density of pits is high alternate with regions with lower intensity/local density.

After having shown that correlation is certainly present among pits, a natural question is whether this correlation is pinned to specific crystallographic directions on the substrate. To clarify this point, we investigated the angular dependence of the spatial structure of pits by evaluating the angle-resolved pair distribution function $g(\mathbf{r}, \theta)$. As we have seen, the integrated conditional density measures the average global density up to a distance $r$ around a given pit. Instead, the function $g(\mathbf{r}, \theta)$ focuses on the density exactly at a distance $r$ and at an angle $\theta$ with respect to the reference pit. Operationally, we chose a reference pit $i$ and a circle of radius $r$ around it and, then, counted the number of pits within $r$ and $r+\Delta r$ which fell inside the annular sector centered at $\theta$ and with a width $\Delta\theta$. To avoid edge effects, the set of reference pits was chosen so that the circle of interest was always completely contained within the border of the image. We first apply the analysis to the image in Fig. 7(a) where the reference pits used for the evaluation of $g(\mathbf{r}, \theta)$ are highlighted with a different color. The result is shown in Fig. 7(b). Interestingly, the pair distribution function is markedly anisotropic: Very pronounced peaks at the scale of the first shell of neighbors are evidenced along the <110> directions, whereas the peak structure is smoothed along the <100> directions. This is visually clear from two cross-sectional profiles taken at a relative angle of 45° and superimposed to the color map of the $g(\mathbf{r}, \theta)$. The non-isotropic local arrangement of the pits is directly visualized in Fig. 7(c) where we plot the spatial distribution of nearest-neighbor distances (SDNN) for the image in Fig. 7(a). For each reference pit, the plot shows the position of the nearest neighbor, with the color scale representing the local density of nearest neighbors. It can be seen that the highest density of nearest neighbors is indeed along the <110> directions.

It is instructive to repeat the analysis above on an image where a longer-range correlation is visually clear, due to the formation of walls of pits [Fig. 7(d)]. In the image, we notice three main walls along the [110] direction. It is interesting to note that, as observed in most cases, the axis of the wall coincides with the direction of one edge of the pits [Fig. 5(b)]. Along the wall axis, the pair distribution function, reported in Fig. 7(e), shows a sharp nearest-neighbor peaks, as well as a clearly oscillating shape that decays only at large values of $r$ and suggests a correlation at long range. At 90° from the wall axis, these long-range oscillations are not present but we notice a broader feature at large distance which is consistent with the average separation between the adjacent walls. From Fig. 7(d), it is also evident that the pits forming the walls are closer to each other, in agreement with the corresponding SDNN map shown in Fig. 7(f) which indicates a larger number of nearest-neighbor along the axis of the walls.

**IV. CONCLUSIONS**

We studied annealing-induced pit-like defects in Ge wafers which are formed during thermal cycles to $T > 750$ °C, i.e. in a temperature range typically used for preparation of the Ge surface before epitaxial growth of overlayers. Morphologically, the pits are analogous to those observed by preferential chemically etching of dislocations. We suggest that the pits are indeed formed by the preferential desorption of material within the strained region around a dislocation pile-up. Despite pits with distinct shapes being observed both on Ge(001) and Ge(111), in the case of the former their density was much higher in line with the reduced stability of the (001) face with respect to the (111) surface of Ge. The spatial distribution of the pits is non-Poissonian and shows evidence for correlation compatible with the mutual interactions between dislocations.

## ACKNOLWDGEMENTS

We gratefully thank E. Placidi, F. Nanni and P. Prosposito for the support with AFM, SEM and optical measurements, respectively. L.P., M.D. and L.D. acknowledge the European Union's Horizon 2020 research and innovation programme under grant agreement No. 766719- FLASH project.

**Table 1 – Size parameters of the pits.**

| Substrate | T (°C) | Density ($x10^5$ cm$^{-2}$) | Area (µm$^2$) | Perimeter (µm) |
|---|---|---|---|---|
| Ge(001) | 770 | 0.8 | $2.5 \pm 0.1$ | $0.4 \pm 0.1$ |
| Ge(001) | 810 | 1.2 | $7.2 \pm 0.3$ | $6.7 \pm 0.6$ |
| Ge(001) | 850 | 6.0 | $11 \pm 3$ | $12 \pm 4$ |
| Ge(111) | 790 | 0.3 | $8.1 \pm 0.3$ | $12 \pm 4$ |
| Ge(111) | 820 | 0.4 | $13 \pm 4$ | $15 \pm 3$ |
| Ge(111) | 830 | 0.4 | $16 \pm 4$ | $18 \pm 5$ |

**FIGURES**

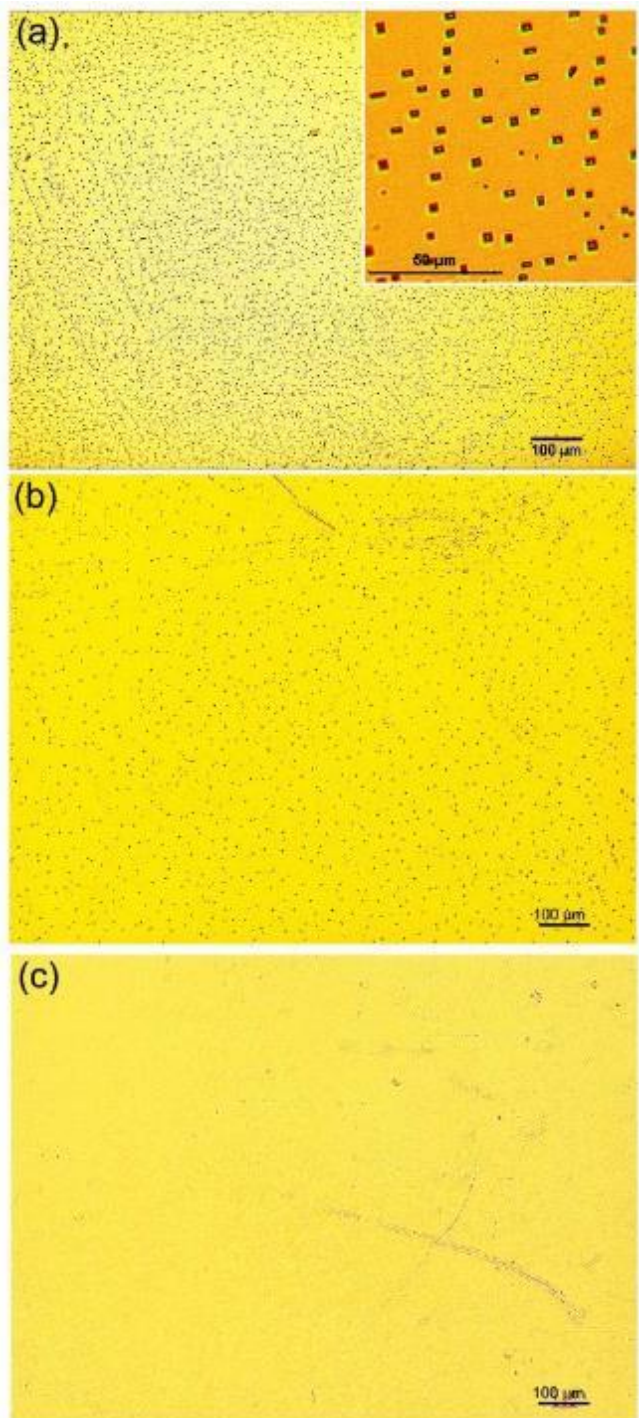

FIG. 1. Optical microscopy images of Ge(001) substrates: (a) after 7 annealing cycles to $T_{max}$= 850 °C, (b) after 7 annealing cycles to $T_{max}$= 810 °C, (c) before any annealing treatment. In the inset of panel (a), the pits are imaged at a higher magnification. The edges of the pits are aligned along the <110> directions.

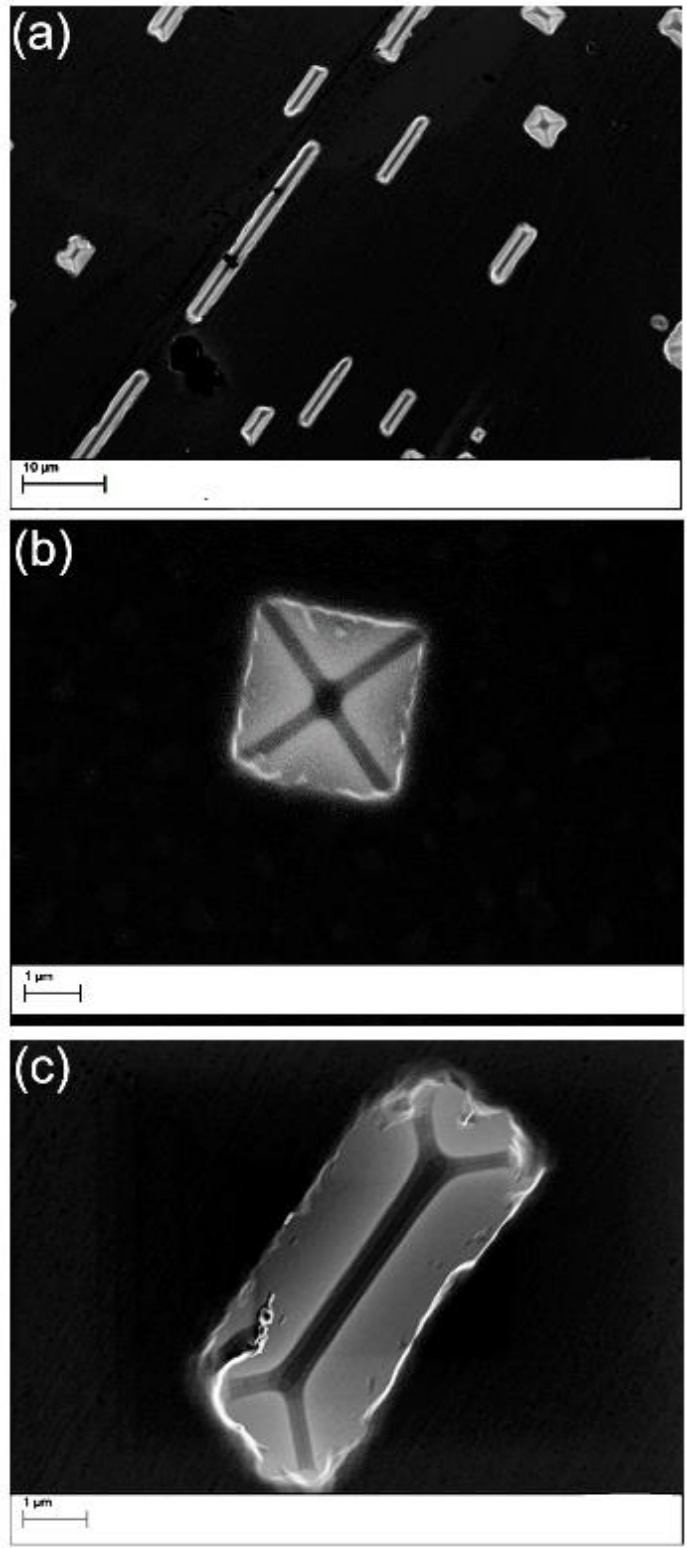

FIG. 2. SEM image of the pits observed on Ge(001) after 7 annealing cycles to $T_{max}$= 850 °C. The edges of the pits are aligned along the <110> directions.

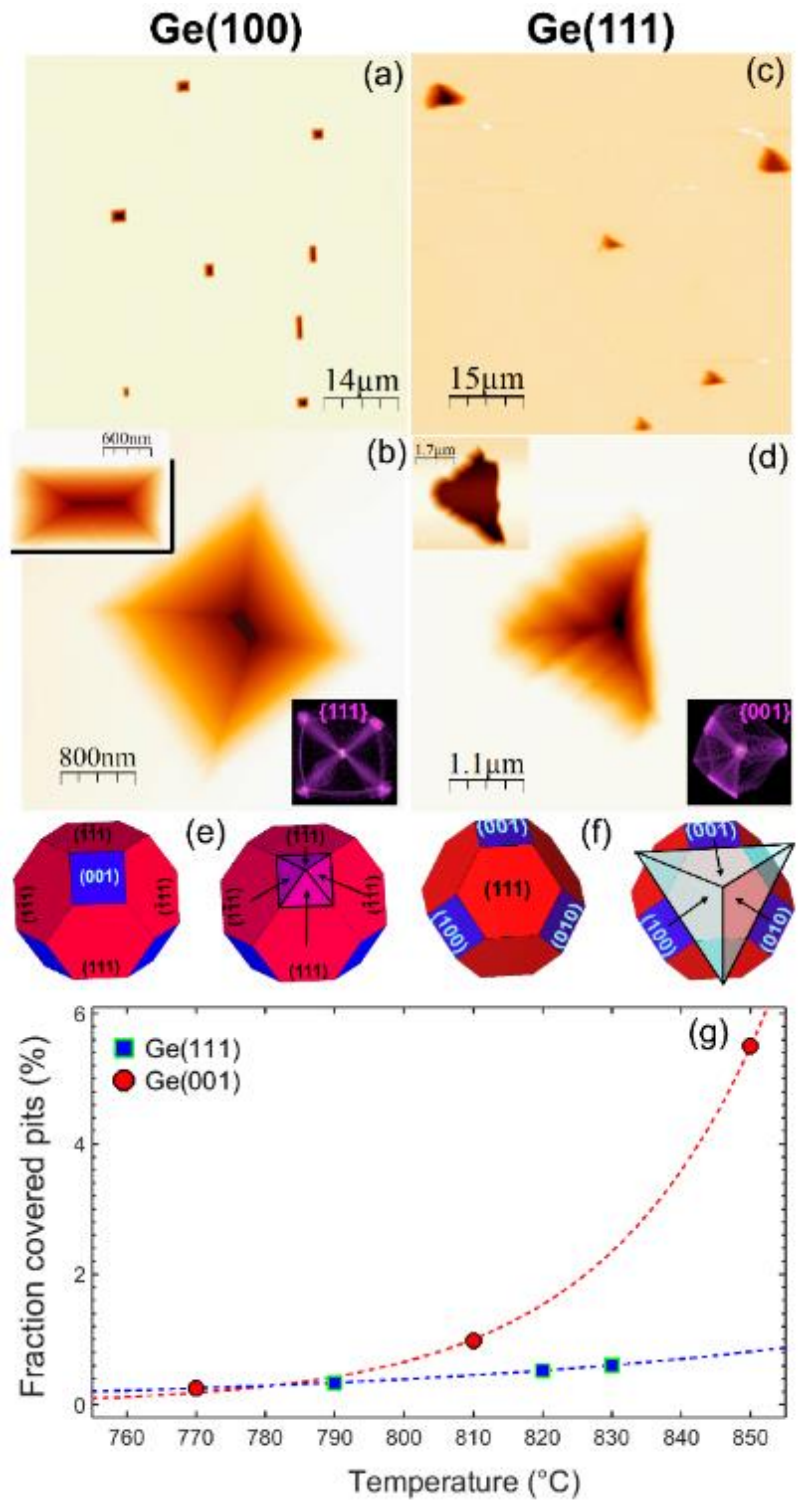

FIG. 3. (a-d) AFM images of pits on Ge(001) and Ge(111). The facet plots of pits on the Ge(001) and Ge(111) surfaces are shown, respectively, in panels (b) and (d). (e, f) Schematics showing the pit faceting on Ge(001) and Ge(111) according to the Wulff shape construction. Panel (g) shows the fraction of the area covered by pits as a function of temperature for the two crystal orientations.

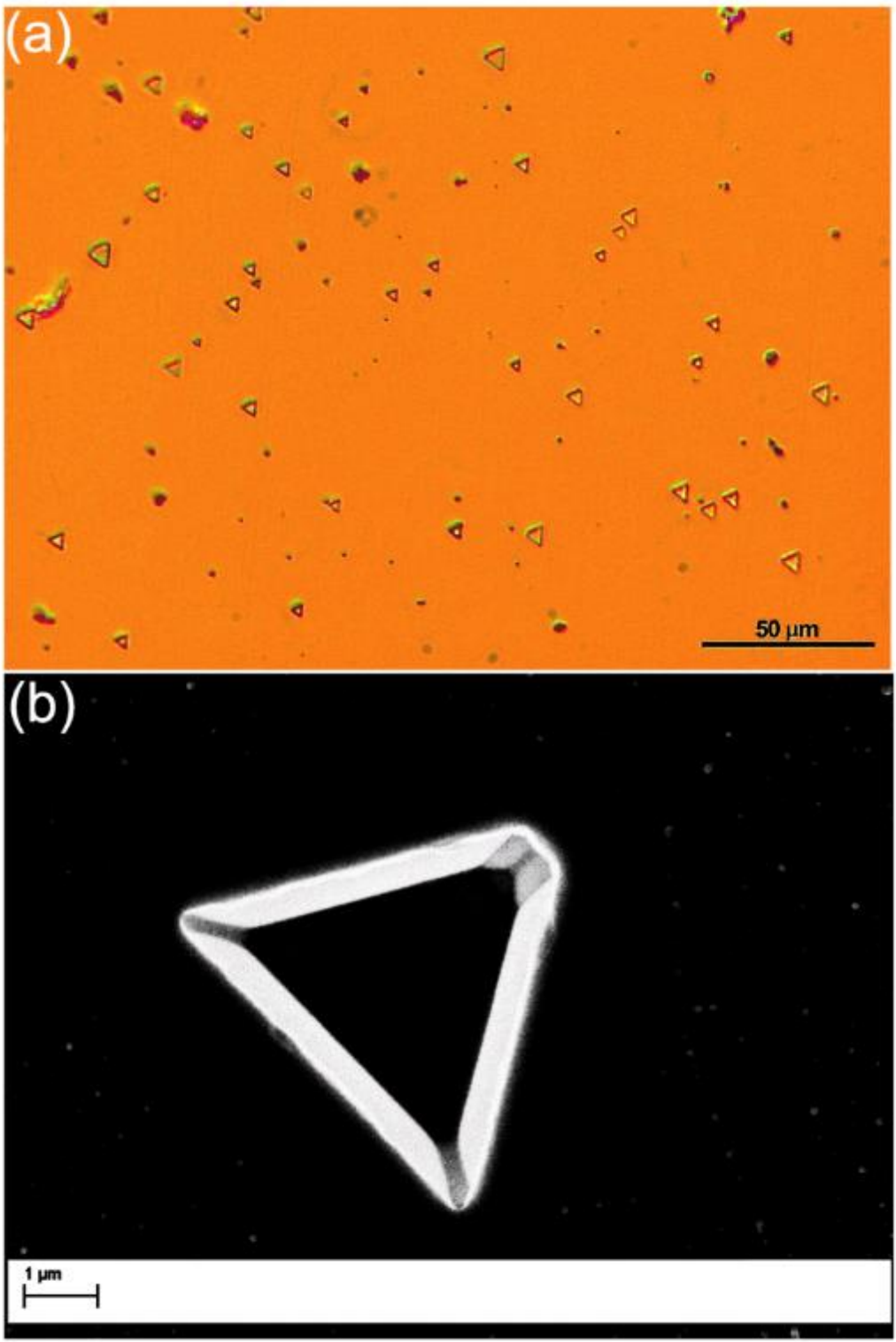

FIG. 4. Triangular pits on the Ge(111) surface after 7 annealing cycles to 820 °C: (a) optical microscopy image; (b) SEM image.

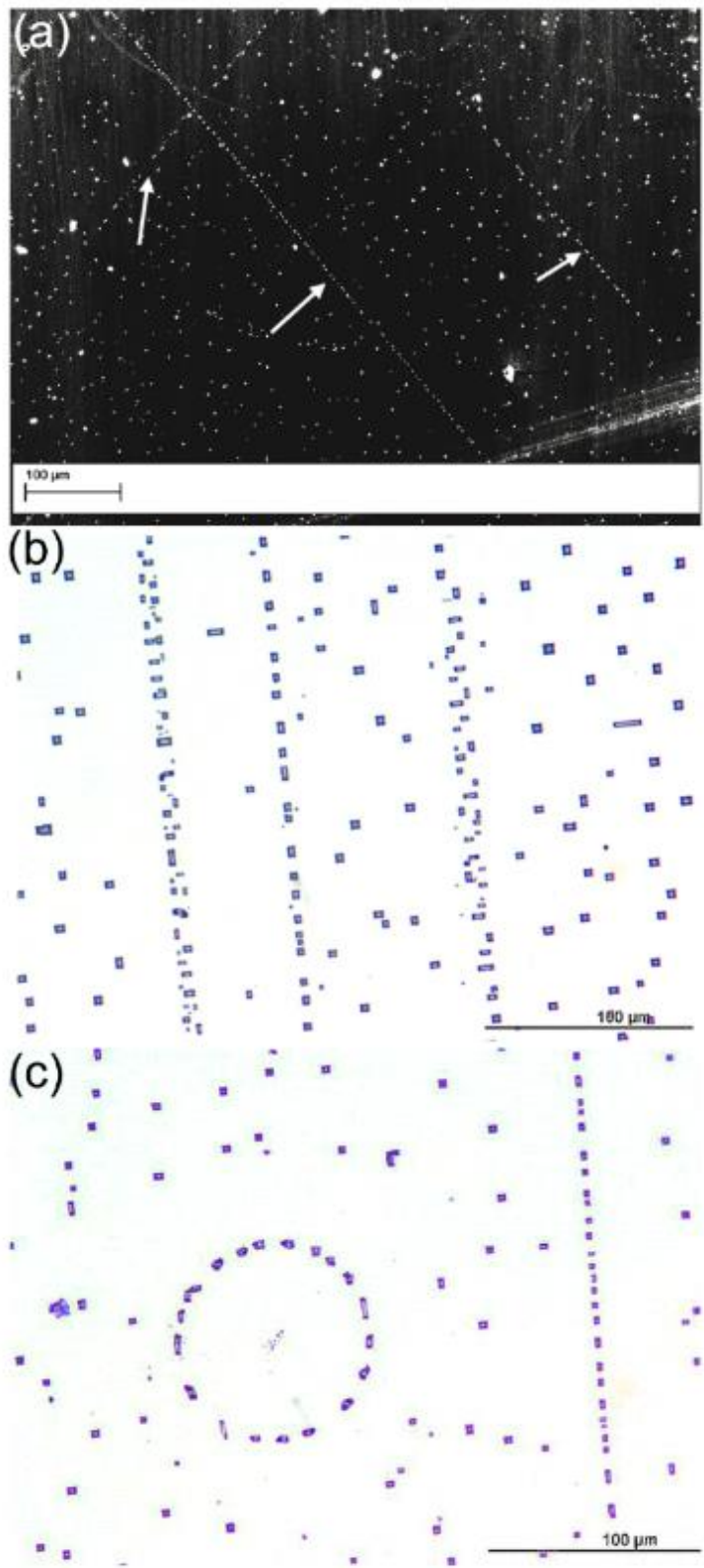

FIG. 5. Optical microscopy images of pit pattern observed on Ge(001). In panel (a), walls of pits are indicated by arrows.

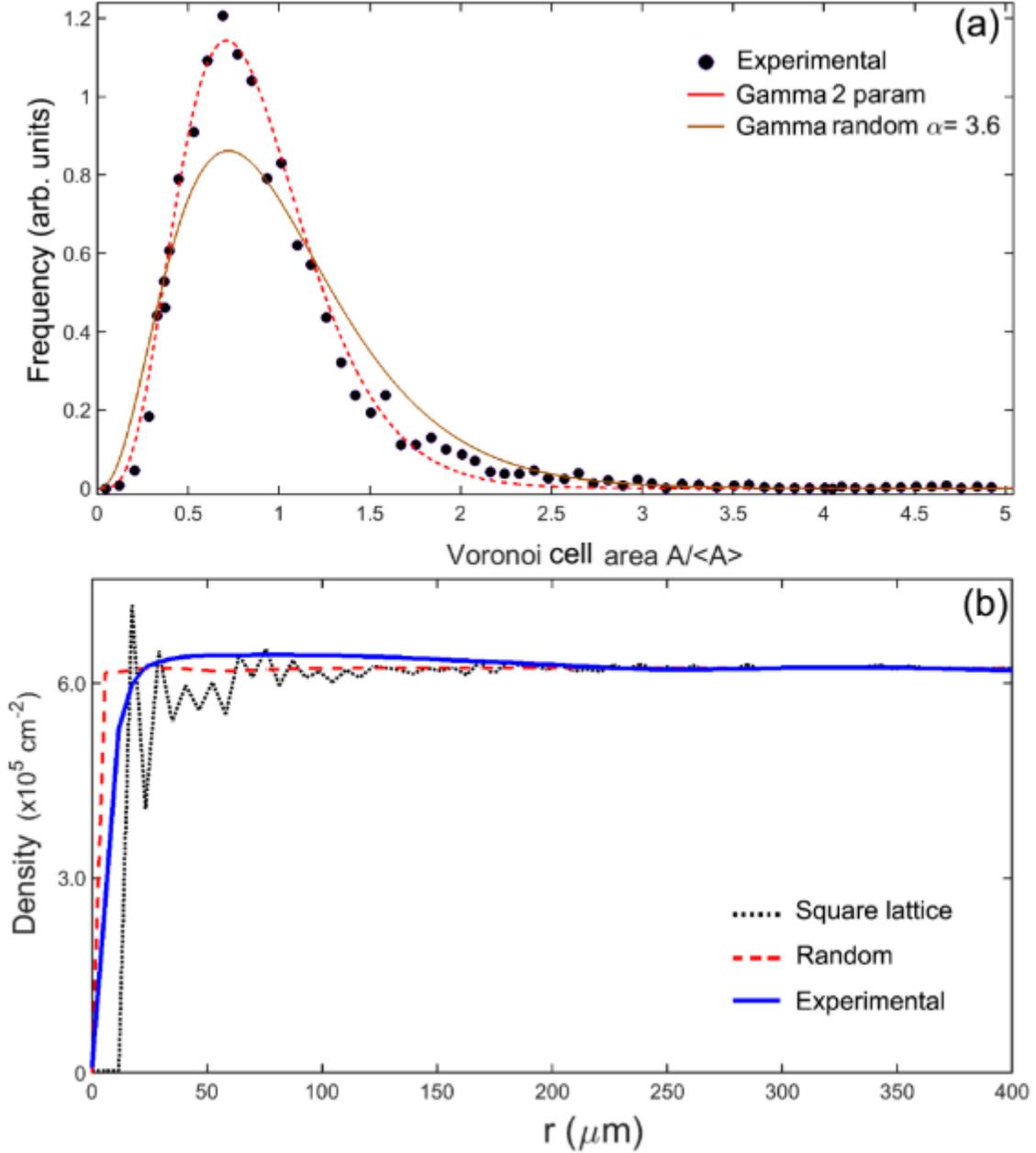

FIG. 6. (a) Distribution of Voronoi cell areas on Ge(001). The dots are the experimental data, the continuous line is the Gamma distribution with α= 3.6 which is expected for a random configuration of points and the dashed line is a fit to the data with a two-parameter Gamma function, $\gamma_{\alpha,\beta}(x) = \beta^{\alpha}\Gamma(\alpha)^{-1} x^{\alpha-1} e^{-\beta x}$ with α= 5.32 and β= 6.1, used as a guide for the eye. (b) Pit density as a function of the distance from the center of mass of each pit averaged over the ensemble of pits. The continuous line is the experimental curve obtained from optical microscopy images. The behavior expected for a random and an ordered square-array of pits with the same asymptotic density are displayed as dashed and dotted lines, respectively.

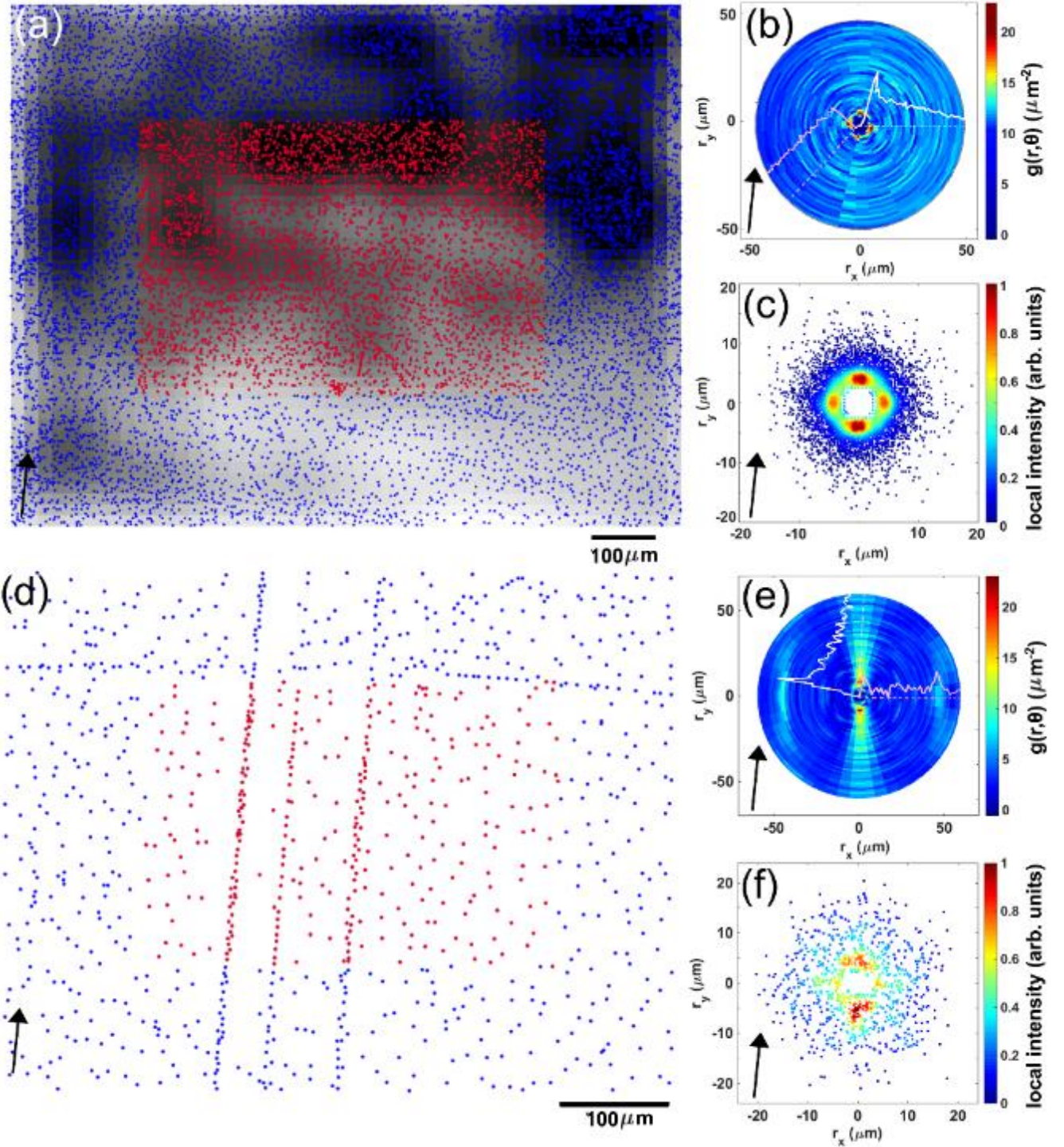

FIG. 7. (a) Binarized image obtained applying the thresholding algorithm to the optical microscopy data of pits on the Ge(001) surface at $T=$ 850 °C. The reference pits used as acceptable centers for the statistical analysis are marked by a different color. (b) Angle-resolved pair distribution function calculated from the image in (a). (c) Spatial distribution of nearest-neighbor distances of panel (a). (d) Binarized image of pits showing the formation of walls on the Ge(001) surface at $T=$ 850 °C. (e, f) Pair distribution function and spatial distribution of nearest-neighbor distances obtained from panel (d). The arrows indicate the [110] direction.